\documentclass[aps,prl,twocolumn,epsf,superscriptaddress]{revtex4}

\usepackage{amsfonts}
\usepackage{amsmath}
\usepackage{amssymb,amsbsy}
\usepackage{bm}
\usepackage{makeidx}
\usepackage{latexsym}
\usepackage{graphicx}
\usepackage{epsfig}
\usepackage{subfigure}
\usepackage{dsfont}
\usepackage{mathcomp}
\usepackage{color}
\usepackage{mathrsfs}
\usepackage{epsf}
\usepackage{pstricks}
\newcommand{\beq}{\begin{eqnarray}}
\newcommand{\eeq}{\end{eqnarray}}
\newcommand{\be}{\begin{equation}}
\newcommand{\ee}{\end{equation}}

\newcommand{\gapp}{\mathrel{\raise.3ex\hbox{$>$}\mkern-14mu
              \lower0.6ex\hbox{$\sim$}}}
\newcommand{\lapp}{\mathrel{\raise.3ex\hbox{$<$}\mkern-14mu
              \lower0.6ex\hbox{$\sim$}}}


\begin{document}
\title{Voids as Alternatives to Dark Energy and the Propagation of Gamma Rays through the Universe.}
\author{Arnaud DeLavallaz}

\affiliation{Physics, Kings College London, Strand, London WC2R 2LS, UK}
\author{Malcolm Fairbairn}

\affiliation{Physics, Kings College London, Strand, London WC2R 2LS, UK}

\begin{abstract}
\noindent
We test the opacity of a void Universe to TeV energy gamma rays having obtained the extra-galactic background light in that Universe using a simple model and the observed constraints on the star formation rate history.  We find that the void Universe has significantly more opacity than a $\Lambda$CDM Universe, putting it at odds with observations of BL-Lac objects.  We argue that while this method of distinguishing between the two cosmologies contains uncertainties, it circumvents any debates over fine-tuning.
\end{abstract}

\maketitle

\begin{figure}[t]
\includegraphics[scale=0.35]{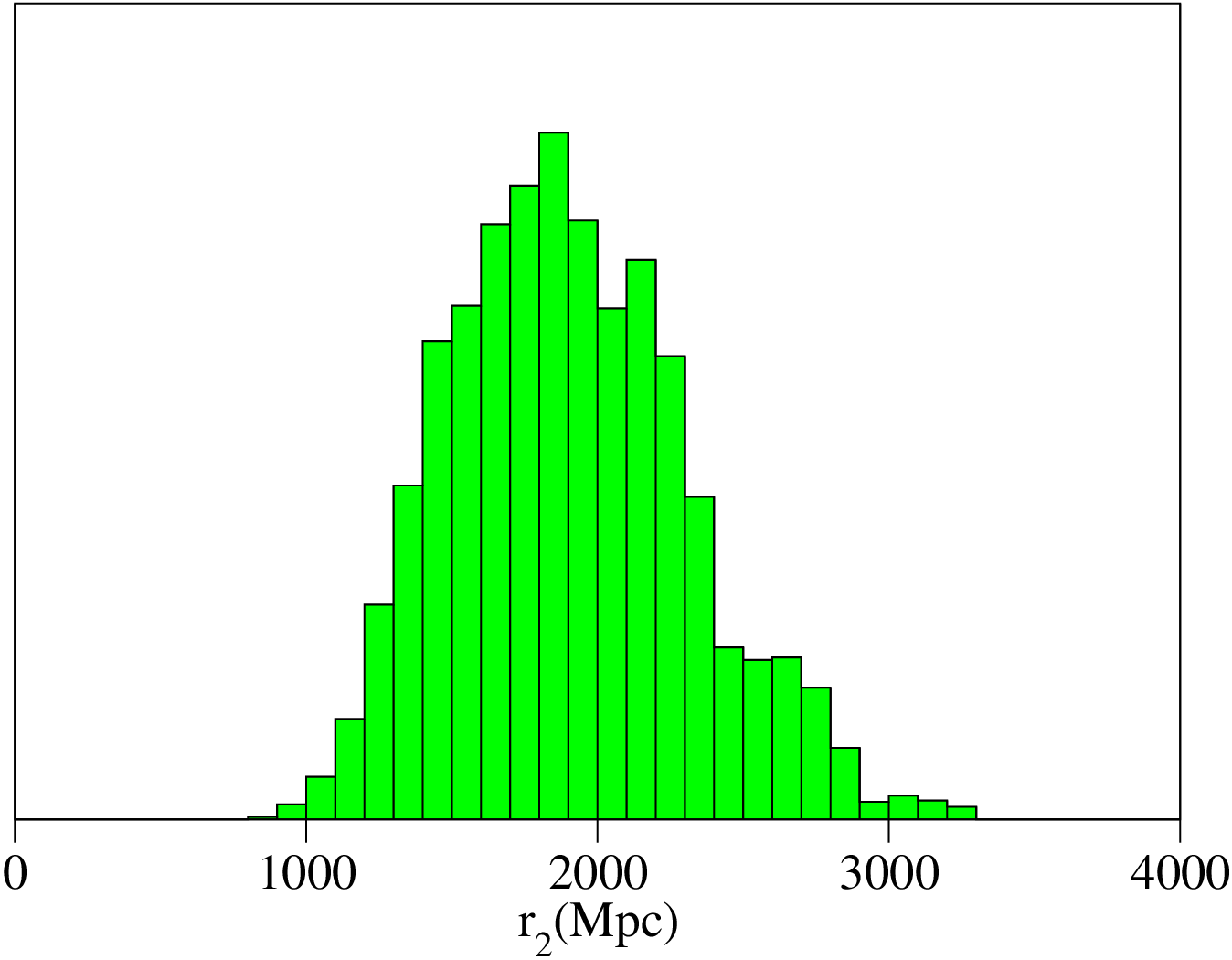}
\caption{Distribution of the size of the void, $r_2$ of equation (\ref{kostov}) necessary to produce voids which fit the supernova data well.{\label{r2plot}}}
\end{figure}

Observations of type 1a supernovae suggest that the expansion of the Universe is accelerating \cite{snorig}.  This fact is usually explained by the presence of some component of energy density which has an equation of state $w\le -1/3$.  This dark energy would be very small compared to the density of matter at early times, becoming much larger relative to matter at late times, which leads to a coincidence problem:  The energy density of the dark energy must be quite finely tuned in order to explain the shape of the supernova data and the density of dark energy and matter would be equal at some redshift $z\sim 0.3$, in other words, very close to today in the history of the Universe when viewed in units of redshift.
It can be argued that our living so close to a particular special time in the history of the Universe goes against the spirit of the cosmological principle.  This, combined with our complete lack of inspiration from particle physics as to what the dark energy might be, has led many researchers to seek alternatives.  However, increasing quantities of data from supernovae \cite{union2}, galaxy surveys \cite{sdss} and the CMB \cite{wmap} only seem to make the dark energy hypothesis stronger - if we can trust the data, it is extremely difficult to live without dark energy.  One notable class of examples of alternatives to dark energy are modified gravity theories, where we change the assumption that the Universe is described by general relativity on large scales \cite{modifiedgravity}.  Another class are void models \cite{voidorig}, which are the models we will focus on in this work.  
These models keep the assumption of general relativity but remove the assumption of isotropy and homogeneity.  The idea is that we are located close to the centre of a very large underdense region compared to the average density of the Universe.  Our local underdense region would then act like a $k=-1$ FRW Universe, the non-zero local $\Omega_k$ contributing the local expansion rate such that if we look outwards and try to recreate the apparent expansion history of the Universe making the assumption it is homogeneous then it appears to be accelerating.  The metric is therefore no longer the isotropic and homogeneous Robertson-Walker metric but if we assume that the void is spherical, we can use the Lema\^{i}tre-Tolman-Bondi (LTB) metric
\begin{equation}
ds^{2}=-c^2 dt^{2}+S^{2}(r,t)dr^{2}+R^{2}(r,t)(d\theta^{2}+sin^{2}\theta d\phi^{2})\label{eq:LTB0}\end{equation}
where the comoving coordinates are $(r,\theta,\phi)$ and time is $t$. For matter without pressure, Einstein's equations imply the following constraints:\begin{equation}
S^{2}(r,t)=\frac{R^{\prime2}(r,t)}{1+2E(r)},\label{eq:LTB1}\end{equation}
\begin{equation}
\frac{1}{2}\dot{R}^{2}-\frac{GM(r)}{R(r,t)}=E(r),\label{eq:LTB2}\end{equation}
\begin{equation}
4\pi\rho(r,t)=\frac{M^{\prime}(r)}{R^{\prime}(r,t)R^{2}(r,t)},\label{eq:LTB3}\end{equation}
here a dot stands for partial derivative with respect to $t$ and
a prime with respect to $r$; $\rho(r,t)$ is the energy density of
the matter and $G$ is Newton's constant. To
specify the model we intend to use, we have to define the two arbitrary
functions $E(r)$, corresponding to the spatial curvature, and $M(r)$,
which is simply the mass integrated within a comoving radial coordinate
$r$:\[
M(r)=4\pi\int_{0}^{r}\rho(r,t)R^{2}R^{\prime}dr.\]
Both of these functions are completely defined therefore by the choice of an initial density profile $\rho(r,t_{\text{LTB}})$,
where $t_{\text{LTB}}=t_{\text{LTB}}(r)$ refers to the beginning
of the LTB evolution and is set to a constant for simplicity. We then
have to choose $E(r)$ in order to match the flat FRW model at the
boundary of the hole. This choice also ensures that the average density inside the hole equals the one outside, so that an observer situated outside the void would not be aware, \textit{locally}, of the presence of the hole.
By choosing the size of the void and the shape and magnitude of the Underdensity we can now fit the supernova data without the presence of dark energy.  Each LTB model we fit to the data is parametrised by the initial density profile $\rho(r,t_{\text{LTB}})$ defined at some early time $t_{LTB}$.  The two functions $E(r)$ and $M(r)$ based on  this profile. The functional form of the initial density profile is based on Kostov's parametrisation \cite{Kostov2009} where the density profile is defined as follows:\begin{multline}
\rho(r,t_{0})=\bar{\rho}(t_{0})\times\\
\left\{ A_{1}+A_{2}tanh\left[\alpha\left(r-r_{1}\right)\right]-A_{3}tanh\left[\beta\left(r-r_{2}\right)\right]\right\}\label{kostov} ,\end{multline}
for $r<r_{h}$, where $r_{h}$ is the radius of the void; and $\rho(r,t_{0})=\bar{\rho}(t_{0})$
for $r\ge r_{h}$. Given $(r_{1},r_{2})$, the values of the coefficients
$(A_{1},A_{2},A_{3})$ and $(\alpha,\beta)$ are chosen so that $\rho(r_{h},t_{0})=\bar{\rho}(t_{0})$
and the integrated mass inside the hole $M(r_{h})=4\pi\int_{0}^{r_{h}}\rho(r,t_{0})r^{2}dr=\frac{4}{3}\pi\bar{\rho}(t_{0})r_{h}^{3}$.   There are therefore three independent parameters in each fit.
\begin{figure}
\begin{center}
\includegraphics[scale=0.35]{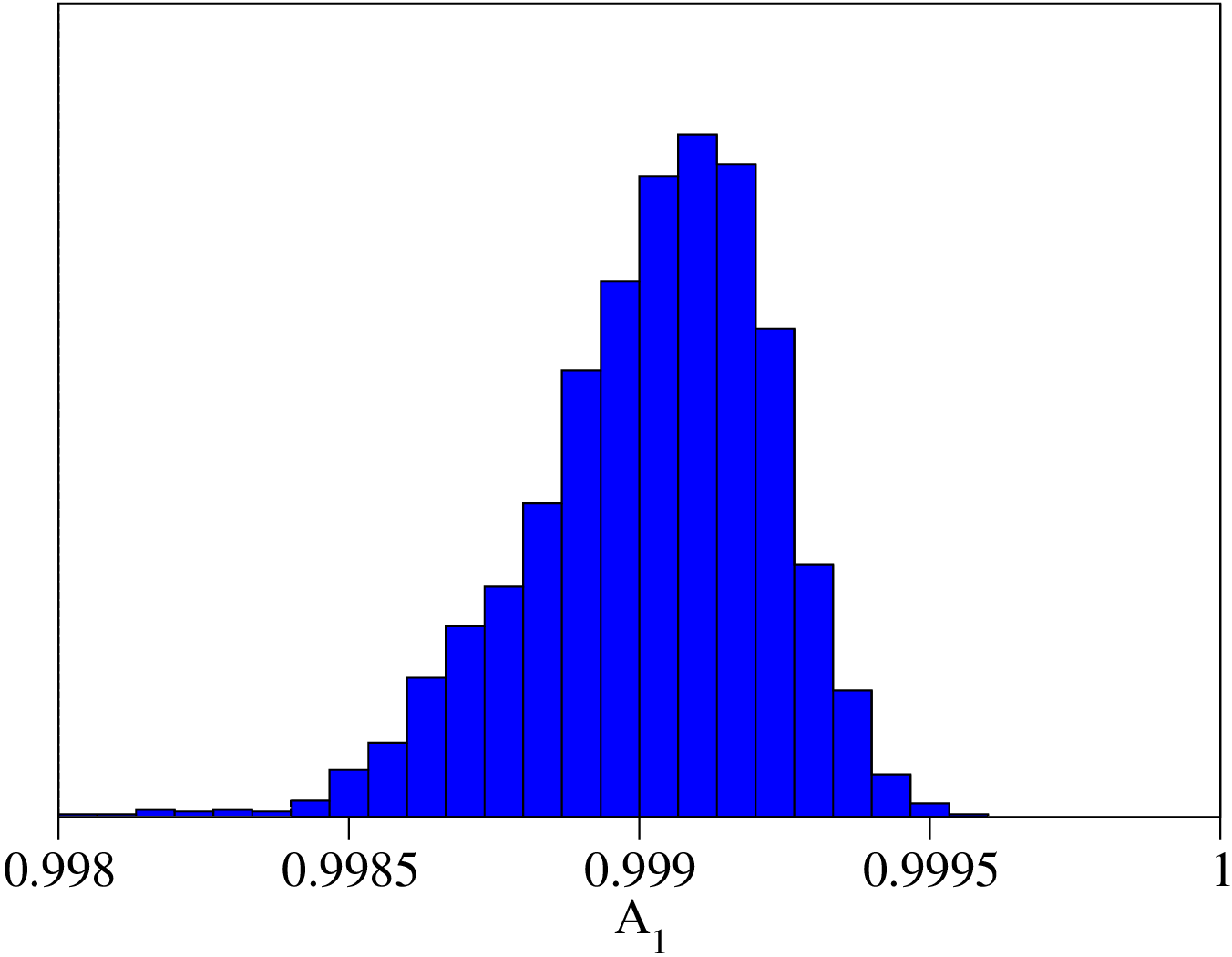}
\caption{Distribution of initial density contrast, $A_1$ of equation (\ref{kostov}) within the void at $z\sim 1100$ necessary to produce voids which fit the supernova luminosity data.\label{a1plot}}
\end{center}
\end{figure}
We perform fits to the supernova data by varying these initial conditions, the position of the solar system relative to the centre of the void, as well as the value of $H_0$ in the homogeneous part of the Universe Outside the void where it is constant.  We do this using a Metropolis Hastings MCMC algorithm, the results of which for the size of the hole and the initial density contrast are shown in figures \ref{a1plot} and \ref{r2plot} respectively.
It is well known that in order to fit the supernova data, the situation needs to be fine-tuned, in particular we need to be very close to the centre of the void for the observations to fit \cite{notari}.  If we look at the size of the volume in which the observer has to be relative to the overall volume of the underdensity, it is typically less than 1\% which obviously constitutes a fine tuning of our position in the void.  
Also, we know that voids do exist in the Universe we live in, but their size is typically only 15-30 Mpc in radius \cite{realvoids}.  While such voids may have a small effect upon the reconstruction of the dark energy equation of state \cite{voidpaper1} they cannot possibly explain the apparent acceleration present in the supernova data.  A quick look at figure \ref{r2plot} shows us that any void required to explain the supernova data would have to be two orders of magnitude larger.  The power spectrum of primordial perturbations must have non-trivial features in it in order for such a large void on a large scale to exist with a non-vanishing probability \cite{powerspectrum}.

Taking a void model seriously would therefore require dropping many of the assumptions that we normally make when we do cosmology.  However, so long as the void model can fit the data and we only have philosophical objections to it, it is not really clear whether we should disregard it outright even though it has more free parameters.  Unlike any of the alternatives, void models only require general relativity and matter, so it can be argued that they should be given more room before we neglect them outright.  
In this work, we aim to show that there is another tool which can be employed to distinguish between the two scenarios, namely the extragalactic background light.  The continual life cycle of stars throughout the history of the Universe creates a background radiation in addition to the CMB.  This Extragalactic Background (star)Light or EBL therefore creates a spectrum of light from the ultra-violet to the far end of the infrared band, up until the point at which the CMB starts to dominate.  Because this background light is made of starlight, and we are surrounded by stars, it is quite difficult to see.  Nevertheless, there are lower limits on its magnitude across the whole range of frequencies due to galaxy counts and positive detections from FIRAS in the infrared. 
 We model the EBL following closely the simplified approach of Finke et al. \cite{finke} with a particular numerical implementation which lends itself to different expansion histories.  Let us outline our way of implementing the equations, the photon density of a blackbody is
\be
n_*(\epsilon,m,t_*)=\frac{dN}{d\epsilon dV}=\frac{8\pi}{\lambda_C^3}\frac{\epsilon^2}{\epsilon\left[\epsilon/\Theta(m,t_*)\right]-1}
\ee
where $\epsilon=h\nu/m_ec^2$ is the dimensionless photon energy in units of electron mass.  The dimensionless temperature $\Theta(m,t_*)=k_BT(m,t_*)/m_ec^2$ where $T(m,t_*)$ is the temperature of a star of mass $m$ and age $t_*$.  We neglect the effect of metallicity variation over time.   We want to know how many photons of energy $\epsilon$ are emitted per unit time from a star of mass $m$ and age $t_*$
\be
\dot{N}(\epsilon,m,t_*)=\frac{dN}{d\epsilon dt}=4\pi R(m,t_*)^2cn_*(\epsilon,m,t_*)
\ee
where $R(m,t_*)$ is the radius of a star of mass $m$ and age $t_*$ and $c$ is the speed of light.  We obtain the temperature $\Theta$ and radius $R$ of a star of mass $m$ and age $t_*$ we follow Finke et al. in using the results from the pa
per by Eggleton, Fitchett and Tout \cite{egg} as well as the corrections outlined in the Finke work \cite{finke}.  For a population of stars of age $t_*$ but with different masses, we need a weighted integral over the range of masses to include the effects of all the different stars.  We therefore define the specific photon generation rate (per unit solar mass)
\be
\tilde{\dot{N}}(\epsilon,t_*)=\frac{dN}{d\epsilon dt}=f(\epsilon)\int_{m_{min}}^{m_{max}}\xi(m)\dot{N}(\epsilon,m,t_*)dm
\ee
where $\xi(m)$ is the initial mass function (IMF) of the stellar population, which we assume doesn't vary over time.  We assume a Salpeter 1955 IMF for the stars that are produced, namely $\xi(m)\propto m^{-2.35}$ across a range of $m_{min}=0.1 M_\odot\le m\le m_{max}=100 M_\odot$.  Here we have also introduced the effects of dimming, the parameter $f(\epsilon)$ is the fraction of photons which do not make it out of galaxies from the blue end of the spectrum due to dust. 
The energy emitted per solar mass of a particular stellar population created at time $t_{form}$ over its lifetime up to the time of the observer (e.g. today) $t_0$ is
\begin{eqnarray}
&&\tilde{j}_{*part}(\epsilon,t_0,t_{form})=\nonumber\\
&&m_e c^2\epsilon\int_{t_{form}}^{t_0}\frac{\tilde{\dot{N}}[(1+z(t))\epsilon,t-t_{form}]}{(1+z(t))}dt
\end{eqnarray}
The dust re-radiates the absorbed energy thermally and modelling the multiple dust grain populations is essential to recreate the $z=0$ spectrum, but not to calculate the opacity of the Universe which mainly depends upon the blue end of the spectrum for TeV scale photons.
\begin{figure}[t]
\includegraphics[scale=0.35]{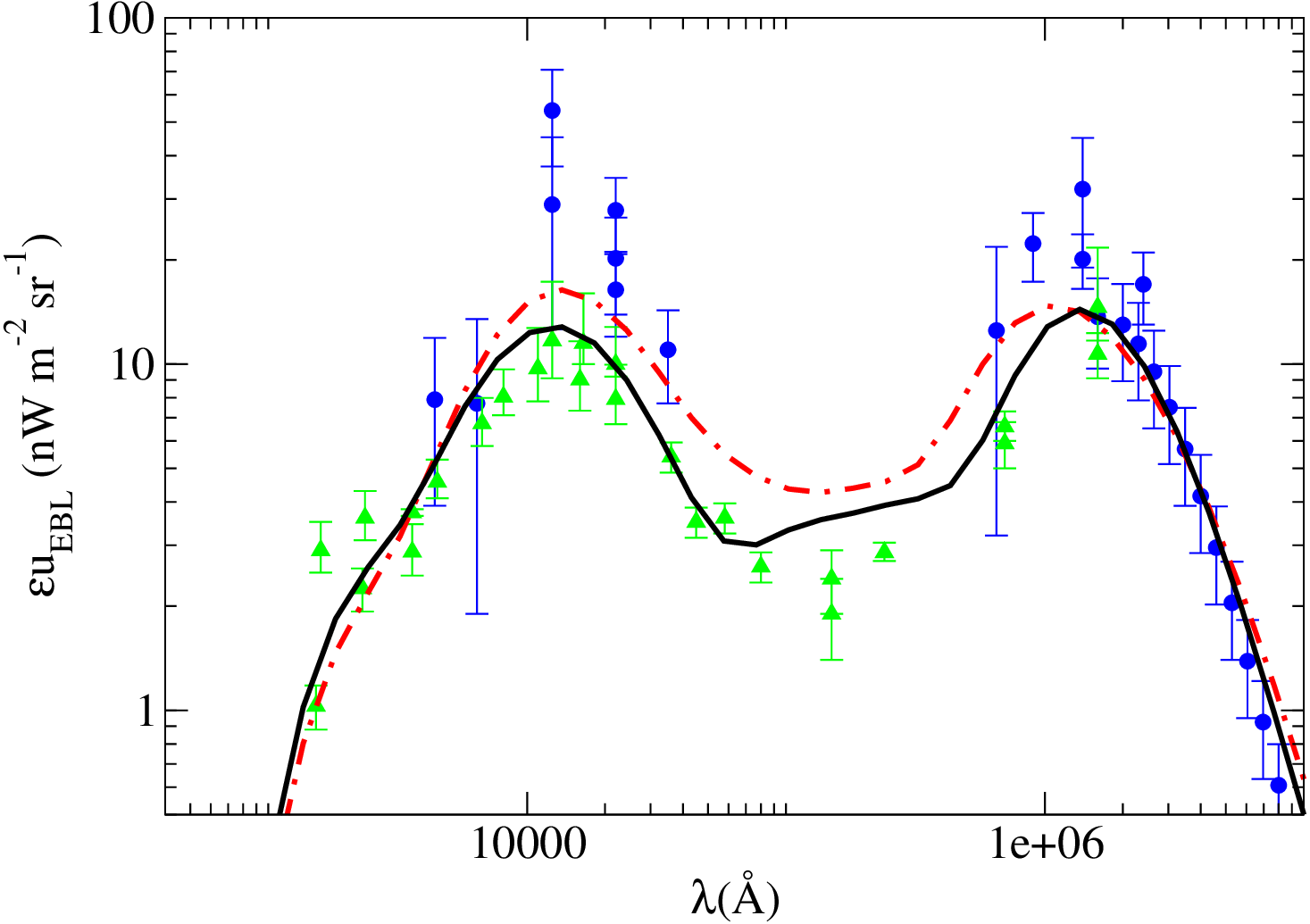}
\caption{The EBL spectrum vs. the data for the two situations of the best fit to the supernova data in the void model (red dot-dashed curve) and the $\Lambda$CDM case (solid black curve).  This is plotted against actual data where the EBL light has been detected (blue circles) and other constraints which only place lower limits (green triangles).  References for the data points are set out in the caption of figure 4 of \cite{primack}.  The blue stellar bulge of these spectra are not fitted to this data but arise from the fit of equation (\ref{sfreq}) to the star formation history rate.{\label{spectrum}}}
\end{figure}
The overall spectrum of energy density at $t_0$ is finally given by
\begin{eqnarray}
&&\epsilon u_{ebl}(\epsilon,t_0)=\nonumber\\
&&\epsilon \int_{t_{start}}^{t_0}\dot{\rho}[z(t)]\left\{\tilde{j}_{*part}(\epsilon,t_0,t)+\tilde{j}_{d}^{part}(\epsilon,t_0,t)\right\}dt
\end{eqnarray}
where $\dot{\rho}(z)$ is the star formation rate in units of solar masses per unit time per unit volume.  To calculate the opacity of the Universe to photons we use
\begin{eqnarray}
\tau(t)=\int_t^{t_0} \int_0^\infty &&c\sigma(2\epsilon E_\gamma (1+z(t'))^2)\nonumber\\
&&\times\frac{u_{EBL}(\epsilon,t')(1+z(t'))^3}{\epsilon m_e c^2}d\epsilon dt
\end{eqnarray}
where $E_\gamma$ is the energy of the gamma ray at $t=t_0$ and $\sigma(y)$ is the angle averaged cross section for electron positron pair production.  The attenuation in photons (i.e. the probability of them getting to $z=0$ when emitted at an earlier time $t$) is then given by $P=\exp(-\tau(t))$.
We adopt the following commonly used functional form for the star formation rate
\begin{equation}
\dot{\rho}_*=\frac{a+bz}{1+(z/c)^d}
\label{sfreq}
\end{equation}
and we fit the values of $a,b,c,d$ to fit the observational constraints on the star formation rate $\dot{\rho}_*(t)$ taken from reference \cite{sfr}.  We choose one of our best fit void cosmologies and compare it with $\Lambda$CDM. For the case of the void cosmology, we apply the appropriate rescaling to the $\dot{\rho}_*$ data outlined in \cite{sfr} to make up for the difference between the void cosmology and the $\Lambda$CDM cosmology.  Both for the void cosmology and the $\Lambda$CDM cosmology, one obtains in this way a very satisfactory $z=0$ spectrum upon tuning the temperature for the re-emission of absorbed light by dust, as can be seen in figure \ref{spectrum}
We have therefore introduced a new data set (the star formation rate history) and shown that it cannot on its own discriminate between the void cosmologies and $\Lambda$CDM if we test it using the $z=0$ spectrum (note the EBL spectrum and the star formation rate data are not independent of each other).  However by moving to a new cosmology, we have changed the relationship between redshift and time, so the physical photon density at any given redshift is different which is critical for the propagation of TeV energy gamma rays.
The photons in the EBL with energies up to and beyond an eV can make the Universe opaque to TeV photons due to electron-positron pair production.  The increased density of EBL photons in void Universes reduces the mean free path of TeV photons and we should not see them coming as abundantly from objects at very large distances.
Recent observations from both the HESS and MAGIC Cerenkov gamma ray detectors have shown that TeV gamma rays are arriving from galaxies at truly cosmological distances, (e.g. 3C 279 at a redshift of $z=0.536$) \cite{hess,magic}. 
\begin{figure}[t]
\begin{center}
\includegraphics[scale=0.35]{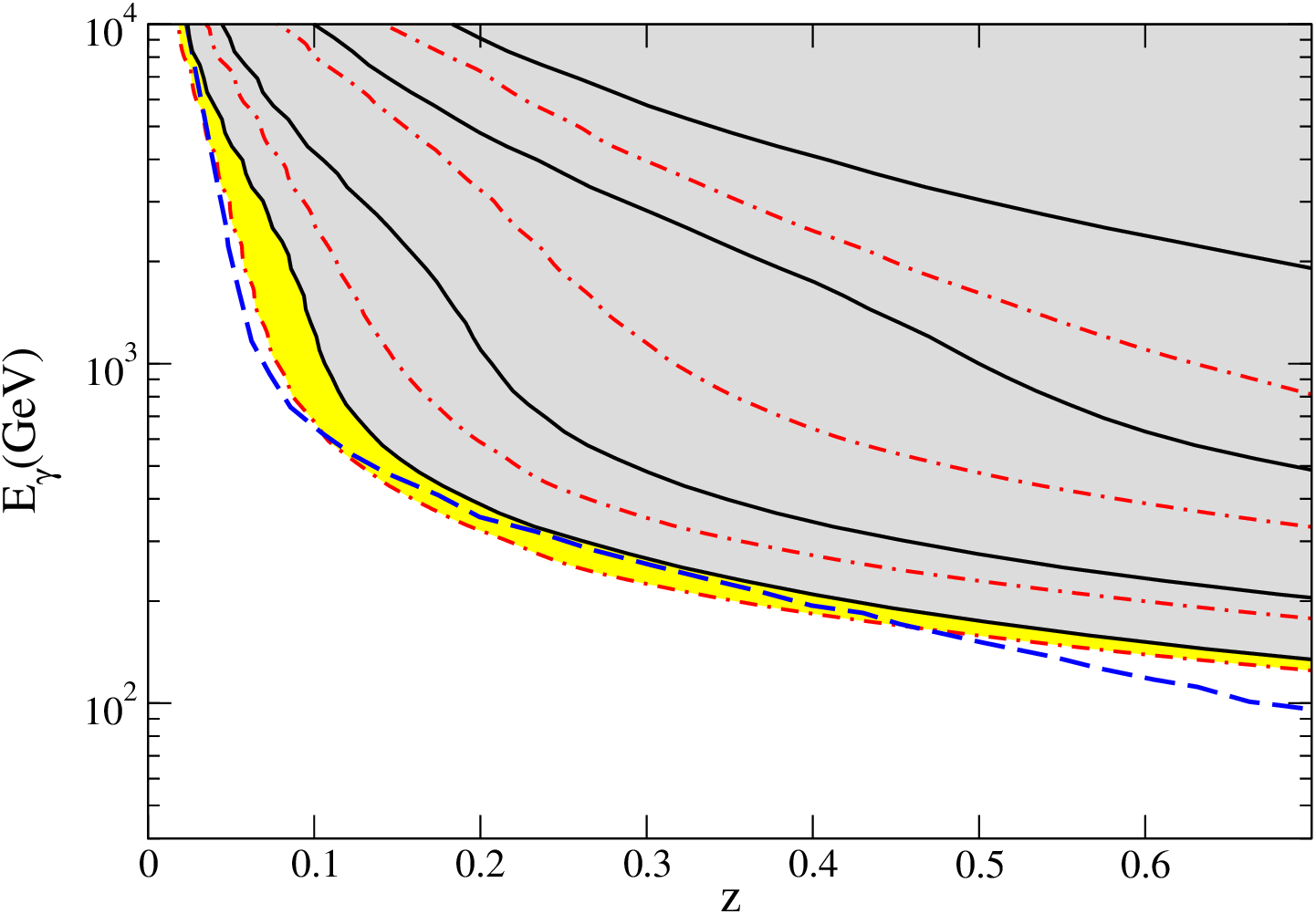}
\caption{Results of our code for opacity $\tau$ for the two different situations of a void Universe and a $\Lambda$CDM Universe.  The solid Black lines correspond to the energy dependent redshift at which the optical depth $\tau=1,2,5,10$ moving from left to right.  The dot-dashed red lines represent the same optical depths for the void cosmology.  The blue dashed line is the constraint from observations \cite{magic}.}\label{compare}
\end{center}
\end{figure}
In figure \ref{compare} we show our results vs. the constraint from the HESS telescope on the optical depth of the Universe to gamma rays.  It is clear that our $\Lambda$CDM background light does not create problems for the propagation of photons, although it is very close to coming into disagreement with the data.  The void cosmology is starting to be in conflict with the data between redshifts of $z=0.15$ and $z=0.4$  In particular, this puts it at odds with the derived limits obtained by the MAGIC Cerenkov telescope from the spectra of the BL-Lac objects 1ES 1101-232 and 1ES 0347-121.  This seems to put the void scenario at odds with the data.
It is important to point out that here we have assumed a particular IMF which does not vary as a function of redshift, neglected the evolution of metallicity and chosen a particular uniform dust reddening.  However, it is clear from figure \ref{compare} that there is a significant difference between the void scenario and the $\Lambda$CDM scenario which we expect to remain when all these other parameters are altered.  More data with regards to the star formation history of the Universe, further observations of the EBL and observations from the Cerenkov Telescope array will test this conclusion and in principle allow us to test other alternatives to $\Lambda$CDM such as modified gravity and more general dark energy models.

\end{document}